\documentclass[conference]{IEEEtran}
 \usepackage{fancyhdr}

\usepackage{graphicx, float, listings, array}
\usepackage{tikz, url}
\usepackage{dblfloatfix}

\usepackage{makecell}
\usepackage{amsmath,amssymb,amsfonts}
\usepackage{algorithmic}
\usepackage{graphicx}
\usepackage{textcomp}
\usepackage{xcolor}
\def\BibTeX{{\rm B\kern-.05em{\sc i\kern-.025em b}\kern-.08em
    T\kern-.1667em\lower.7ex\hbox{E}\kern-.125emX}}

\usetikzlibrary{arrows.meta, positioning, shapes, fit}

\usepackage[most]{tcolorbox}

\newtcolorbox{qbox}{
enhanced,
boxrule=0pt,frame hidden,
borderline west={4pt}{0pt}{gray!75!black},
colback=gray!10!white,
sharp corners
}

\newtcolorbox{finding}{
enhanced,
boxrule=0pt,frame hidden,
borderline west={4pt}{0pt}{violet!75!black},
colback=violet!10!white,
sharp corners
}

\newcolumntype{C}[1]{>{\centering\arraybackslash}p{#1}}

\usepackage[skip=2pt]{caption} 

\usepackage[table]{xcolor}
\definecolor{resultfocus}{RGB}{245,248,232}

\begin{document}

\title{BitIR: Cross-Architecture Fault Injection for Resilience Analysis of Heterogeneous GPU Applications
}
\author{
  \IEEEauthorblockN{Maisy Dunlavy\IEEEauthorrefmark{1},
                    Michael E. Papka \IEEEauthorrefmark{2}\IEEEauthorrefmark{1},
                    Zhiling Lan\IEEEauthorrefmark{1}\IEEEauthorrefmark{2}}
  \IEEEauthorblockA{\IEEEauthorrefmark{1}University of Illinois Chicago, USA \quad
                    \IEEEauthorrefmark{2}Argonne National Laboratory, USA\\
                    \{mdunl6, papka, zlan\}@uic.edu}
}
\maketitle

\thispagestyle{fancy}
\lhead{}
\rhead{}
\chead{}
\lfoot{\footnotesize{
To Appear at SC26, November 15-20, 2026, Chicago, Illinois, USA
}}
\rfoot{}
\cfoot{}
\renewcommand{\headrulewidth}{0pt}
\renewcommand{\footrulewidth}{0pt}

\begin{abstract}
Modern GPU-based HPC systems rely on heterogeneous vendor stacks, yet resilience studies are largely limited to single architectures, leaving it unclear how faults behave across different GPU backends. 
We present \emph{BitIR}, a cross-architecture fault injection framework that injects deterministic single-bit faults at the LLVM IR level, ensuring semantically equivalent perturbations prior to backend lowering and enabling direct cross-vendor comparison across NVIDIA, Intel, and AMD GPUs.
We evaluate BitIR on three production supercomputers -- Polaris (ALCF, NVIDIA A100), Aurora (ALCF, Intel  GPU Max 1550), and Frontier (OLCF, AMD Instinct MI250X) -- representing the full spectrum of current leadership-class GPU architectures. 
Using representative heterogeneous benchmarks, we conduct large-scale injection campaigns across all three systems and classify outcomes into masked results, silent data corruptions (SDCs), and failures. 
Our results show that identical faults produce markedly different behaviors across vendors: Intel most often masks faults but exhibits more hangs when faults escape masking, NVIDIA exposes more detectable hard failures, and AMD alternates between SDC-dominant and failure-dominant behavior depending on the benchmark and fault site. These findings demonstrate that resilience is not backend-invariant, underscoring the need for backend-aware and workload-aware fault mitigation.
\end{abstract}

\section{Introduction}

Graphics Processing Units (GPUs) are now central to modern high-performance computing (HPC) and AI systems. As these platforms continue to scale, they become increasingly vulnerable to transient faults caused by radiation, voltage noise, and timing variation~\cite{MukherjeeHPCA2005,Oliveira2016}. Although hardware protection mechanisms mitigate some of these effects, many faults still propagate to software-visible state and can ultimately alter application behavior. In heterogeneous applications, where GPUs execute the dominant computational kernels, even a single bit perturbation can lead to silent corruption, abnormal termination, or stalled execution ~\cite{Tan2011,TiwariHPCA2015}.

Understanding how GPU applications behave under transient faults is, therefore, an important resilience problem. This is especially true for scientific and data-intensive workloads, where incorrect outputs may be difficult to detect and where GPU kernels can fail silently without OS-level signals or reliable log evidence. A useful resilience study must therefore characterize faults at the application level, based on software-visible outcomes rather than only low-level hardware events ~\cite{ReisTACO2005,Kalra2020}.

\emph{A key open question} is whether the same logical fault (i.e., a semantically equivalent perturbation applied at the same program point) produces the same outcome across different GPU backends. None of the existing literature has been able to address this directly because they are typically confined to a single vendor stack or a single abstraction level. Hardware and microarchitecture level approaches provide detailed insight into many complex GPU resilience issues~\cite{Tan2011,Vallero2018}, but they are tightly coupled to particular architectures and do not support direct cross-platform comparison. Software-based approaches improve portability and allow for the potential of cross-platform comparison~\cite{Fang2014,SASSIFI2017,NVBitFI2021}, but many operate on backend-specific representations such as assembly or vendor binaries, making it difficult to define an equivalent fault across different GPU ecosystems.

This gap motivates a cross-architecture fault-injection methodology 
grounded in a common, software-visible abstraction. We introduce 
\textbf{BitIR} (\emph{Bit-flip Intermediate Representation}), a unified framework for resilience analysis across heterogeneous GPU platforms that 
injects deterministic single-bit faults at the LLVM intermediate 
representation (IR) level before backend-specific lowering. By operating 
on the shared IR rather than on vendor-compiled binaries, BitIR defines 
faults in terms of program semantics rather than backend-specific 
encodings, which allows the perturbation to be truly 
equivalent across NVIDIA (CUDA), Intel (SYCL), and AMD (HIP) compilation 
flows. BitIR targets three classes of software-visible values: instruction 
results, operands, and pointer-related values. It also supports deterministic site enumeration, aligned worklist construction, backend-specific lowering, 
and output-based fault classification against golden references, all of which enables
large-scale campaigns on real production systems while preserving a shared 
fault model across vendors. 

Unlike many prior GPU fault-injection studies that rely on statistical or randomized sampling of fault locations for tractability~\cite{Fang2014,SASSIFI2017,NVBitFI2021,gpuFI42022}, BitIR exhaustively evaluates the complete aligned injection space for each benchmark. This is particularly important for cross-vendor comparison, where sparse sampling can miss backend-specific divergence concentrated in specific workloads, instruction classes, or site categories.

\begin{figure*}[t]
    \centering
    \includegraphics[width=0.9\linewidth]{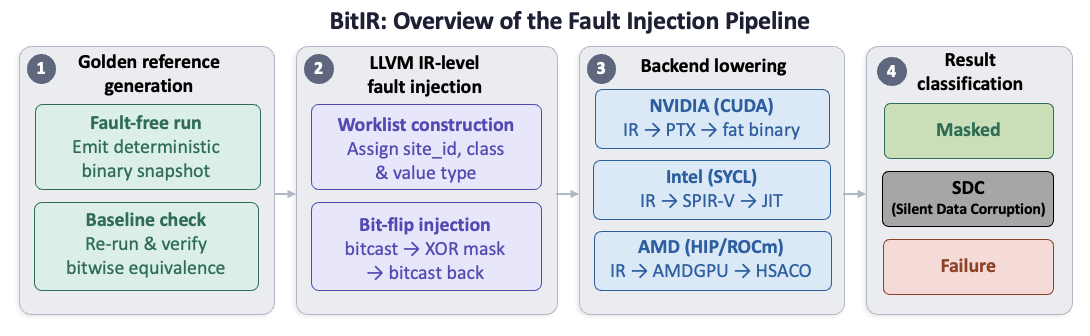}
    \caption{Overview of the BitIR pipeline.}
    \label{fig:overview}
\end{figure*}

Using BitIR, we conduct large-scale injection campaigns on representative heterogeneous benchmarks across three production supercomputers: Polaris (ALCF, NVIDIA A100), Aurora (ALCF, Intel Data Center GPU Max~1550), and Frontier (OLCF, AMD Instinct MI250X). Our results show that resilience is not backend-invariant and that the same aligned fault often produces drastically different outcomes for each vendor. These differences are especially pronounced for memory-access and pointer-bearing sites and vary substantially by workload, underscoring the need for backend-aware and workload-aware fault mitigation.  
\\

Specifically, this work makes the following contributions:

\begin{itemize}
\vspace{0cm}\item We introduce BitIR, the first LLVM IR-based framework for cross-architecture fault injection across NVIDIA, Intel, and AMD GPUs, featuring deterministic site enumeration, aligned worklist construction, and reproducible campaign execution on 
production supercomputers~(\S\ref{sec:method}).
    \item We conduct an extensive empirical study using BitIR across three leadership-class supercomputers (Polaris, Aurora, and Frontier) on representative heterogeneous benchmarks, comparing execution outcomes under a shared IR-level fault model~(\S\ref{sec:imp}).
    \item We show that resilience differences across GPU backends are strongly 
    shaped by workload structure and injected IR site characteristics, with concrete implications for backend-aware and workload-aware 
fault mitigation~(\S\ref{sec:results}).
    \item We discuss guidelines and recommendations for fault mitigation strategies tailored to the hardware platform on which an application is deployed~(\S\ref{sec:implications}).
    
\end{itemize}

\section{Key Challenges}
\label{sec:challenge}
We now identify the practical challenges that arise when studying transient faults at the application level\footnote{We use ``application-level'' and ``software-visible'' interchangeably to denote behavior observable at the program level (e.g., outputs, execution status, or runtime errors), as opposed to microarchitectural or hardware level effects.} in heterogeneous GPU systems. These challenges stem not only from the nature of fault propagation but also from the difficulty of defining consistent fault models and conducting reproducible experiments across diverse backend software stacks.

\textbf{Challenge 1.}
In GPU applications, fault outcomes are often difficult to classify. A program may complete while producing incorrect outputs, or it may report runtime or benchmark-level errors while still producing outputs matching the golden reference. Consequently, exit status and runtime-level signals alone are insufficient for distinguishing masked outcomes, silent data corruption, failures, and hangs~\cite{ReisTACO2005,Kalra2020}. Reliable classification, therefore, requires application-aware comparison against expected outputs.

\textbf{Challenge 2.}
Cross-platform comparison is difficult because NVIDIA CUDA, Intel SYCL, and AMD HIP rely on distinct compiler toolchains and backend representations~\cite{CUDAProgGuide,IntelOneAPI,AMDHIP}. \footnote{SYCL and HIP are portable programming models and are not restricted to Intel and AMD hardware, respectively. In this work, we associate SYCL with Intel and HIP with AMD because these are the primary vendor-supported programming environments used on the evaluated platforms and serve as the closest practical counterparts to CUDA on NVIDIA.}If faults are injected after backend lowering, the injected perturbation is tied to backend-specific instructions, and the resulting campaigns cannot be compared under a common fault model. This limits the ability to draw meaningful cross-architecture conclusions.

\textbf{Challenge 3.}
Large-scale GPU fault injection is also an execution-management problem. Each injection may require recompilation, device code generation, benchmark execution, output collection, and result classification. Scaling to thousands of fault sites and bit positions, therefore, requires deterministic work partitioning, scheduler-aware execution, restartable campaigns, and reliable aggregation from persistent artifacts. Without these capabilities, large campaigns become difficult to reproduce and validate.

BitIR addresses each challenge directly. Outcomes are classified against golden references rather than exit status alone~(Challenge 1); fault injection is performed before backend lowering, enabling 
semantically aligned perturbations across CUDA, SYCL, and HIP~(Challenge 2); 
and experiments are organized as deterministic, worklist-driven 
campaigns with persistent per-run artifacts, enabling scalable 
and reproducible execution~(Challenge 3).

\section{Methodology}
\label{sec:method}

Given a heterogeneous application that offloads compute-intensive kernels to GPUs, our goal is to systematically characterize how single-bit faults affect application resilience across NVIDIA, Intel, and AMD GPU architectures under a common fault model. To enable fair and consistent cross-platform comparison, our methodology enforces deterministic execution, output-equivalent validation, and precise alignment of injected sites across vendor backends. Figure~\ref{fig:overview} illustrates the overall workflow.

At the core of BitIR is the LLVM intermediate representation (IR)\cite{LLVM-IR}, which serves as a unifying software-visible abstraction prior to backend lowering. LLVM IR is typed, static single assignment (SSA)-based, and widely used in modern CPU and GPU compilation pipelines before vendor-specific code generation. Unlike backend binaries, LLVM IR preserves instruction structure, operand relationships, and type information, allowing faults to be defined in terms of program semantics rather than hardware-specific encodings. Using LLVM IR and its debug metadata, BitIR constructs deterministic site/bit worklists, applies aligned injections across platforms, and classifies outcomes through comparison against golden outputs.

\emph{Cross-architecture equivalence} is essential to the validity of our 
study. We define it as follows: equivalence holds if, for all 
platforms, each fault injection satisfies (i)~\emph{input 
equivalence} --- identical program inputs and output comparison 
criteria; (ii)~\emph{fault equivalence} --- the same bit position 
is perturbed in a semantically equivalent LLVM IR object across all 
backends; and (iii)~\emph{interpretation equivalence} --- results 
are classified and aggregated under a common methodology.

\subsection{Golden Reference Generation}
\label{sec:golden-ref}

For each application, we first establish a \emph{golden reference output} by executing the fault-free benchmark with hooks that emits a deterministic binary snapshot of its final program state. This includes output buffers, result arrays, or other application-specific outputs for each benchmark. We verify that all expected outputs are generated and non-empty, then repeat the execution and compare the resulting snapshots for bitwise equivalence. Additional fault-free baseline executions were performed during campaign validation and consistently reproduced the corresponding golden references.
This verification procedure ensures that we are able to produce stable, deterministic outputs for each benchmark that can be used for output classification throughout injection campaigns.

\subsection{LLVM IR-Level Fault Injection}
\label{sec:IR-FI}

Fault injection is performed at the LLVM IR level under a unified fault model through three steps: site enumeration, worklist construction, and injection execution.

\textit{Site Enumeration.}
We traverse the device-side LLVM IR of each application to identify eligible \emph{injection sites}. A site is defined as any LLVM IR-visible target for which a fault can be introduced. Each site is assigned a deterministic \texttt{site\_id} based on a fixed traversal order, ensuring stability and reproducibility across runs on unchanged IR.
 
Each site is annotated with several attributes.  The
\emph{site class} captures the role of the site in the IR. 
\emph{Result} sites model corruption of a computed value, \emph{operand} 
sites perturb an instruction input, and \emph{pointer} sites corrupt a 
device memory address, typically via a \texttt{getelementptr} chain. 
The \emph{value type} determines the bit width and semantic interpretation of the 
perturbation. \emph{Integer} sites cover indices, counters, control-flow 
conditions, and arithmetic results, \emph{floating-point} sites appear 
in arithmetic kernels, where exponent flips shift values by orders of 
magnitude and mantissa flips introduce bounded relative error, and 
\emph{pointer} sites hold memory addresses whose corruption may redirect 
loads or stores to invalid regions, producing outcomes from silent 
overwrite to access violation. 
Additional attributes, such as opcode and function context
are recorded to support cross-platform site alignment.

\textit{Worklist Construction.}
For each enumerated site, BitIR constructs a \emph{worklist} of \texttt{(site\_id, bit\_index)} pairs, where the bit index ranges over all bit positions of the target value for each site (e.g., 32 positions for \texttt{float} and typically 64 for pointers on modern GPU backends). Each site\/bit pair uniquely identifies a single fault injection experiment. These pairs serve as primary keys in both the worklist and the resulting per-run output artifacts. 

We validate the reproducibility and stability of this methodology by regenerating site inventories and confirming identical IDs, bit widths, and metadata across runs on unchanged IR.

\textit{Injection Execution.}
Given a \texttt{(site\_id, bit\_index)} pair, the injector locates the target IR object, bitcasts it to an integer of matching width, applies an XOR with a one-hot mask at the specified bit position, and bitcasts the result back to its original type, guaranteeing a single-bit flip that preserves type correctness. This mechanism allows faults to be introduced uniformly across different LLVM IR-visible targets while keeping the perturbation semantics reproducible.

BitIR records a shared metadata schema for each candidate site, including target class, opcode, type, bit width, operand index, function context, source location, and a signature ordinal that disambiguates sites sharing the same source location. These attributes are combined into what we refer to as a semantic identifier that characterizes the LLVM IR object independently of any vendor-specific profiling or numbering schemas. 

To maintain cross-platform alignment, sites with semantic identifiers that do not map cleanly across vendor compilation flows are pruned. Only 15 of 1150 total injection sites (1.3\%) were removed from the aligned benchmark set. These sites correspond to backend or compiler-specific helper machinery, including CUDA builtin-variable helpers, shuffle intrinsics, and device-function wrappers introduced by Clang CUDA support headers. These sites represent a negligible fraction of candidate sites and preserves near-complete coverage of the aligned injection space. No injection runs were executed for pruned injection sites.

\subsection{Backend Lowering}
\label{sec:backend}

After IR-level injection, the modified IR is lowered through each vendor's native compilation pipeline into a runnable benchmark. The injected fault remains semantically invariant across platforms with the same aligned IR site/bit perturbed before lowering, but we do not assume identical backend compilation or runtime behavior. 
Instead, any differences that emerge during lowering and execution (e.g., instruction selection, register allocation, memory access lowering, and runtime handling) are treated as part of the backend behavior that BitIR is designed to expose.

\begin{itemize}
\item \emph{NVIDIA CUDA 11.8.0:} The injected IR is lowered to PTX and packaged into a fat binary via the CUDA toolchain. 
\item \emph{Intel oneAPI DPC++/C++ Compiler 2025.2.0 
:} The injected IR is lowered to SPIR-V and reintegrated into the host executable, where it is JIT-compiled to native Gen/Xe device code at runtime. 
\item \emph{AMD ROCm 6.4.2:} The injected IR is lowered through the AMDGPU backend via the HIP/ROCm toolchain and linked into the final device binary.
\end{itemize}

To reduce compilation overhead, unmodified device IR is cached when the corresponding source is unchanged; controlled A/B tests confirm that caching does not affect experimental outcomes. 

\begin{figure}[t]
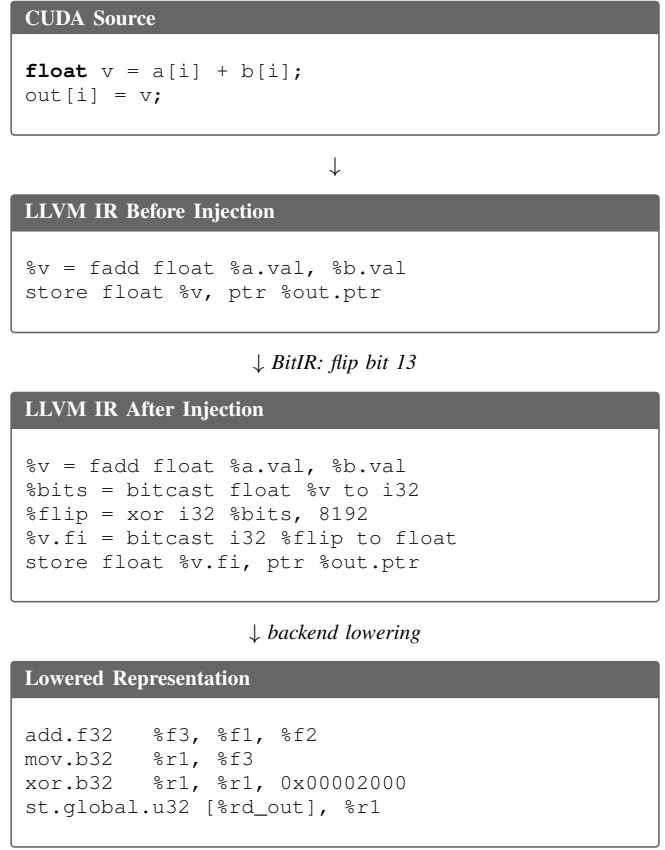

\centering
\begin{minipage}{0.97\columnwidth}
\footnotesize

\begin{tcolorbox}[
    title=\textbf{CUDA Source},
    colback=white,
    colframe=black!60,
    boxrule=0.6pt,
    arc=1pt,
    left=2pt,right=2pt,top=1pt,bottom=1pt
]
\begin{lstlisting}[language=C++,basicstyle=\ttfamily\footnotesize]
float v = a[i] + b[i];
out[i] = v;
\end{lstlisting}
\end{tcolorbox}

\vspace{-0.4em}
\begin{center}
$\downarrow$
\end{center}
\vspace{-0.6em}

\begin{tcolorbox}[
    title=\textbf{LLVM IR Before Injection},
    colback=white,
    colframe=black!60,
    boxrule=0.6pt,
    arc=1pt,
    left=2pt,right=2pt,top=1pt,bottom=1pt
]
\begin{lstlisting}[basicstyle=\ttfamily\footnotesize]
%v = fadd float %a.val, %b.val
store float %v, ptr %out.ptr
\end{lstlisting}
\end{tcolorbox}

\vspace{-0.4em}
\begin{center}
$\downarrow$ \textit{BitIR: flip bit 13}
\end{center}
\vspace{-0.6em}

\begin{tcolorbox}[
    title=\textbf{LLVM IR After Injection},
    colback=white,
    colframe=black!60,
    boxrule=0.6pt,
    arc=1pt,
    left=2pt,right=2pt,top=1pt,bottom=1pt
]
\begin{lstlisting}[basicstyle=\ttfamily\footnotesize]
%v = fadd float %a.val, %b.val
%bits = bitcast float %v to i32
%flip = xor i32 %bits, 8192
%v.fi = bitcast i32 %flip to float
store float %v.fi, ptr %out.ptr
\end{lstlisting}
\end{tcolorbox}

\vspace{-0.4em}
\begin{center}
$\downarrow$ \textit{backend lowering}
\end{center}
\vspace{-0.6em}

\begin{tcolorbox}[
    title=\textbf{Lowered Representation},
    colback=white,
    colframe=black!60,
    boxrule=0.6pt,
    arc=1pt,
    left=2pt,right=2pt,top=1pt,bottom=1pt
]
\begin{lstlisting}[basicstyle=\ttfamily\footnotesize]
add.f32   %f3, %f1, %f2
mov.b32   %r1, %f3
xor.b32   %r1, %r1, 0x00002000
st.global.u32 [%rd_out], %r1
\end{lstlisting}
\end{tcolorbox}

\end{minipage}
\vspace{0.3cm}

\caption{Example of BitIR’s aligned LLVM IR-level fault injection. A fault is specified as a \texttt{(site ID, bit index)} pair before vendor-specific lowering. Although lowering differs across backends, the same pair identifies the same logical perturbation, which survives when the lowered computation depends on the corrupted value.}
\label{fig:lowering-example}
\end{figure}

As a sanity check, we confirmed that injected IR modules successfully propagate through each backend compilation flow and produce observable non-masked outcomes after execution. Figure~\ref{fig:lowering-example} shows a representative example. For conciseness, only the NVIDIA lowering path is shown, although equivalent validation was performed for the Intel and AMD backends.

Across all three platforms, the same IR-level injection mechanism produces SDCs, hard failures, and hangs, demonstrating that the injected perturbations are not systematically eliminated during lowering. Individual \textsc{Masked} outcomes may still arise from backend optimization or later masking, and we do not attempt to distinguish these mechanisms for every injection in this study. 

\subsection{Output Classification}
\label{sec:classification}

Each fault injection produces per-run artifacts (\texttt{siteX\_bitY.out} 
and \texttt{siteX\_bitY.err}) corresponding to a single pair of
\texttt{(site\_id, bit\_index)}. 

\begin{figure}[h]
    \centering
    \includegraphics[width=0.70\linewidth]{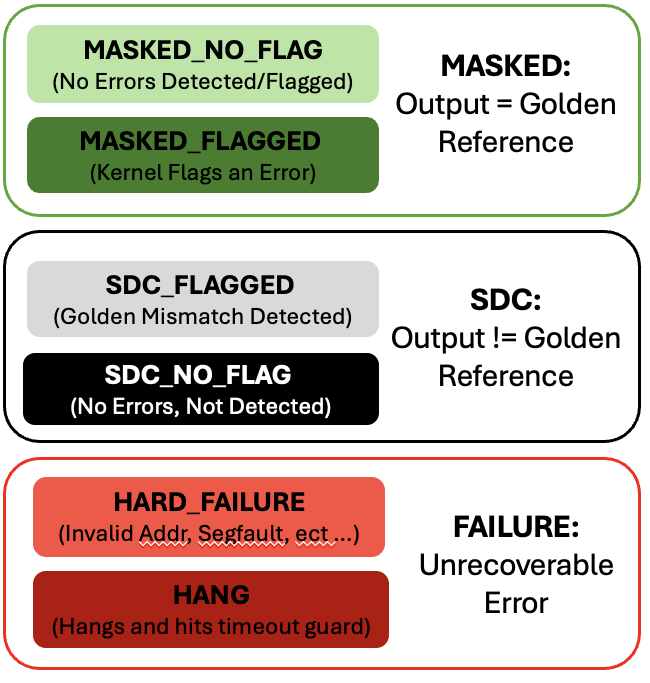}
    \caption{Outcome taxonomy. }
    \label{fig:output}
\end{figure}

The generated output is compared against platform-specific golden references using benchmark-specific comparators and classified as follows (Figure~\ref{fig:output}): (i)~if the output matches within tolerance, the result is \textsc{Masked}; (ii)~if the output differs but the application completes normally and produces a non-empty, well-formed result, it is \textsc{SDC}; and (iii)~if execution terminates abnormally due to a crash, runtime error, or compiler error, it is \textsc{Failure}; executions exceeding a per-run timeout are forcibly terminated and recorded as \textsc{Hang}, a tracked subtype of \textsc{Failure}.

If an output file is missing, the site/bit pair is re-run rather than assumed to be a \textsc{Failure} outcome. Only upon completing twice with no output is it considered to be a \textsc{Failure}. 

BitIR supports multiple output forms, including floating-point arrays, exact binary dumps, and benchmark-specific textual outputs. For floating-point outputs, we perform elementwise comparison using a fixed absolute tolerance of $10^{-5}$ and classify a run as \textsc{Masked} only if all elements satisfy this threshold; otherwise, it is classified as \textsc{SDC}. For our selected floating-point benchmarks, a fixed absolute tolerance was sufficient because fault-free baseline runs compared cleanly against their golden outputs at this threshold, while avoiding the instability of relative error for output values near zero. This threshold was appropriate for our benchmark set, but other benchmarks may require benchmark-specific relative tolerances. For binary outputs, exact bytewise equality is required to be considered \textsc{Masked}.

We implement timeout thresholds to prevent stalled executions from blocking large-scale campaigns while preserving visibility into hang-induced fault behavior. 
Timeout thresholds are set to 200 seconds for all benchmarks, which is substantially higher than the observed fault-free execution times (mean execution time is below 35 seconds for all benchmarks). This ensures that hang classification reflects abnormal execution behavior rather than backend-specific performance differences. 

Aggregate outcome statistics 
are constructed directly from per-run \texttt{*.out} artifacts, 
ensuring reproducibility, campaign restartability, and cross-architecture equivalence.

\begin{table*}[t]
\centering
\small
\begin{tabular}{C{2.2cm} C{1.8cm} C{1.2cm} p{5.3cm} p{5.0cm}}
\hline
System & Peak PFlop/s per Node & Scheduler & CPU Configuration per Node 
& GPU Configuration per Node \\
\hline
Polaris (ALCF) & 0.0388  & PBS 
& 1$\times$AMD EPYC 7543P, 512~GiB DDR4 
& 4$\times$NVIDIA A100, 40~GiB HBM2 \\
Aurora (ALCF) & 0.130 & PBS 
& 2$\times$Intel Xeon Max 9470C, 2$\times$512~GiB DDR5 
& 6$\times$Intel Max~1550, 768~GiB HBM2e \\
Frontier (OLCF) & 0.1912 & Slurm 
& 1$\times$64-core AMD EPYC 7A53, 512~GiB DDR4 
& 4$\times$AMD MI250X, 64~GiB HBM2e \\
\hline
\end{tabular}
\caption{Hardware configuration of the three production 
supercomputers used in this study.}
\label{tab:hardware-platforms}
\end{table*}

\begin{table*}[t]
\centering
\small
\begin{tabular}{c l l c r c c c}
\hline
\textbf{Benchmark} & \textbf{Category} & \textbf{Args} & \textbf{\#Sites} & \textbf{\#Bits} & \textbf{Result} & \textbf{Operand} & \textbf{Pointer} \\
\hline
matrix-rotate   & Math (Index Re-Mapping)             & \verb|[10000, 1]|          & 84  & 10,532 & 27.4\% & 27.7\% & 44.9\% \\
jacobi          & Math (Iterative Propagation)              & None                       & 195 & 20,962 & 31.7\% & 32.3\% & 36.0\% \\
layout          & Classification    & \verb|[1]|                 & 38  & 5,764  & 23.4\% & 23.4\% & 53.2\% \\
dense-embedding & Machine Learning  & \verb|[10000, 8, 1]|       & 132 & 19,564 & 23.4\% & 23.9\% & 52.7\% \\
pathfinder      & Simulation        & \verb|[10000, 1000, 5]| & 154 & 17,831 & 26.0\% & 24.9\% & 49.1\% \\
entropy         & Encoding/Decoding & \verb|[10000, 1024, 1]|    & 184 & 21,306 & 25.2\% & 25.8\% & 49.0\% \\
colorwheel      & Image Processing  & \verb|[10000, 8, 1]|       & 243 & 27,162 & 27.3\% & 31.2\% & 41.5\% \\
randomAccess    & Bandwidth (Memory Access)        & \verb|[1]|                 & 120 & 17,892 & 32.5\% & 32.1\% & 35.4\% \\
\hline
\end{tabular}
\vspace{1em}
\caption{Selected HeCBench benchmarks, execution arguments, and resulting injection scale. \emph{Args} reports the exact command-line values used for each benchmark (argument names, ordering, and benchmark-specific semantics are documented in the corresponding benchmark directory in the HeCBench documentation~\cite{HeCBench}). \emph{Sites} reports the number of eligible LLVM IR-visible injection sites, \emph{Bits} reports the total number of single-bit injections in the benchmark worklist. The final three columns show the distribution of those injections across result, operand, and pointer site classes. }
\label{tab:apps}
\end{table*}

\section{Implementation and Evaluation}
\label{sec:imp}

\subsection{BitIR Implementation}
BitIR is a multi-component LLVM-based GPU fault injection framework built on LLVM 17.0.0 ~\cite{LLVM-IR} and integrated with CUDA, SYCL, and HIP toolchains. It consists of three components: (1) an LLVM pass plugin (\texttt{libfi\_inject.so}) for IR-level fault injection, implemented in C++; (2) backend compilation flows, orchestrated via Bash, that lower injected IR through vendor-specific toolchains into executable binaries; and (3) a Python-based campaign execution layer that manages site inventory generation, worklist construction, cross-platform alignment, result aggregation, and integration with the schedulers.

\subsection{Hardware Platforms}

Experiments were conducted on three production supercomputers: Polaris (ALCF) with NVIDIA A100 GPUs, Aurora (ALCF) with Intel Data Center GPU Max 1550 GPUs, and Frontier (OLCF) with AMD Instinct MI250X GPUs. Table~\ref{tab:hardware-platforms} summarizes the hardware configuration for each system.

\subsection{Benchmarks}

Our experiments use the HeCBench benchmark suite~\cite{HeCBench}, 
which spans a broad range of computational motifs representative of 
modern HPC GPGPU workloads, including numerical kernels, search, image 
processing, and encoding/decoding tasks. HeCBench provides native 
implementations in CUDA, SYCL/DPC++, and HIP for all benchmarks in the suite, making it a prime candidate for any cross architecture study. Most kernels in the suite also include integrated correctness checks and produce deterministic outputs, making them directly comparable against a golden 
reference and simplifying outcome classification. Benchmark parameters 
are chosen to balance execution size, per-injection runtime, and output 
stability across repeated fault-free runs. The injection scale varies across our chosen benchmarks, since each benchmark exposes a different number 
of eligible LLVM IR injection sites with potentially different bit 
widths. Table~\ref{tab:apps} summarizes the selected benchmarks, 
workload categories, runtime parameters, and injection scale.

\subsection{Run-to-Run Stability}

To assess run-to-run stability, we performed repeated executions for a sampled subset of injection points, consisting of 300 site/bit pairs per benchmark spread across each GPU platform evenly. The vast majority of injection outcomes are deterministic across runs. Only 1.42\% of the sampled injections exhibited outcome variability across runs, with pointer-related sites accounting for 76.5\% of unstable cases. This is consistent with the semantics of pointer and address manipulation in low-level parallel programming models, where corrupted addresses may result in implementation-dependent execution behavior~\cite{CUDAProgGuide,IntelOneAPI,AMDHIP}. Given the low rate and structured nature of this run-to-run variability, we conclude that these effects do not materially impact the aggregate trends reported in this work.

\section{Resilience Analysis}
\label{sec:results}

We structure our analysis around five research questions targeting distinct dimensions of cross-architecture fault behavior.

\subsection{Backend Variance}
\label{sec:backend-variance}

\begin{qbox}
\textbf{Question 1.} Do semantically aligned IR-level faults produce consistent outcome distributions across NVIDIA, Intel, and AMD GPU backends, or does resilience vary significantly by platform?
\end{qbox}

\begin{figure*}[t]
    \centering
    \includegraphics[width=\linewidth]{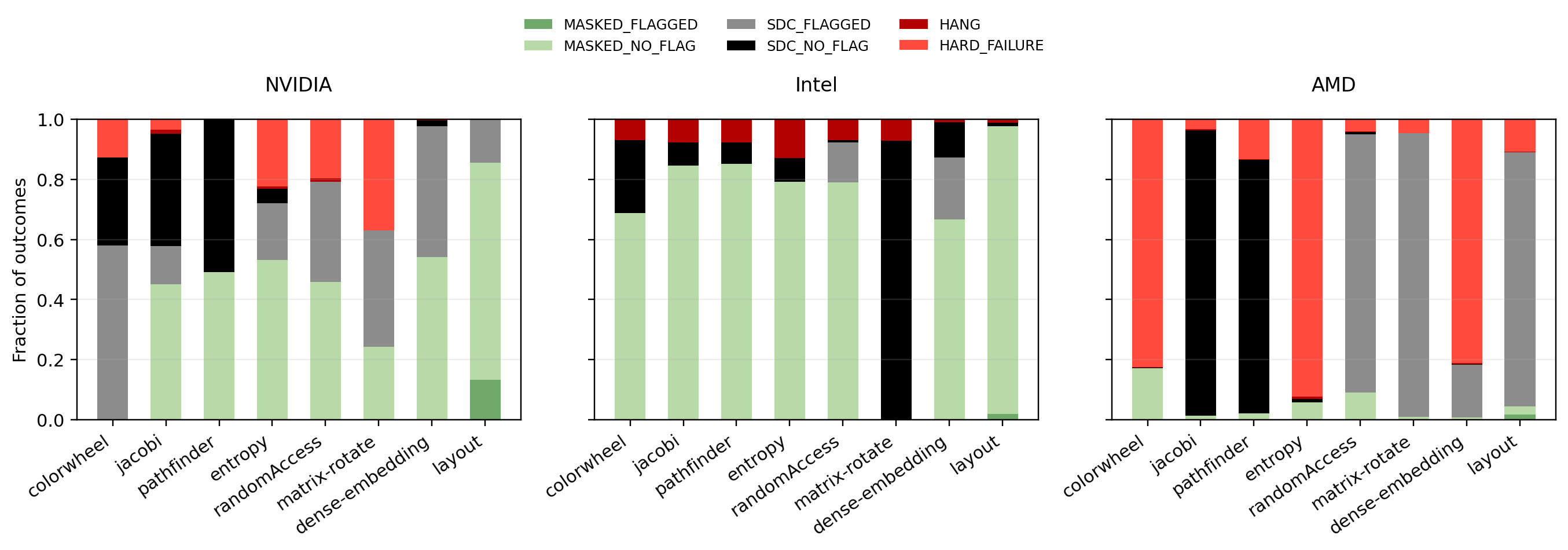}
    \caption{Fine-grained outcome distributions across benchmarks and backends, using the full six-class outcome taxonomy.}
    \label{fig:backend-failure-split}
\end{figure*}

Table~\ref{tab:backend-coarse-summary} summarizes coarse outcome ratios 
across the aligned three-backend benchmark set, and 
Figure~\ref{fig:backend-failure-split} shows the corresponding 
fine-grained breakdown. Together, these views confirm that resilience 
varies substantially by backend.

\begin{table*}[t]
\centering
\small
\setlength{\tabcolsep}{6pt}
\renewcommand{\arraystretch}{1}
\begin{tabular}{p{1.9cm} p{1.1cm} p{1.15cm} p{1.15cm} p{1.15cm} !{\vrule width 0.6pt} p{1.9cm} p{1.1cm} p{1.15cm} p{1.15cm} p{1.15cm}}
\hline
Benchmark & Backend & Masked & SDC & Failure & Benchmark & Backend & Masked & SDC & Failure \\
\hline
\texttt{jacobi} & NVIDIA & 45.1\% & 50.1\% & 4.8\%
& \texttt{matrix-} & NVIDIA & 24.4\% & 38.7\% & 36.9\% \\
& Intel & 84.6\% & 7.6\% & 7.8\%
& \texttt{rotate} & Intel & 0.0\% & 92.8\% & 7.2\% \\
& AMD & 1.3\% & 95.1\% & 3.6\%
& & AMD & 0.9\% & 94.4\% & 4.7\% \\
\hline
\texttt{pathfinder} & NVIDIA & 49.1\% & 50.9\% & 0.0\%
& \texttt{dense-} & NVIDIA & 54.1\% & 45.5\% & 0.4\% \\
& Intel & 85.2\% & 7.1\% & 7.7\%
& \texttt{embedding} & Intel & 66.7\% & 32.4\% & 0.9\% \\
& AMD & 2.1\% & 84.4\% & 13.4\%
& & AMD & 0.7\% & 17.8\% & 81.5\% \\
\hline
\texttt{entropy} & NVIDIA & 53.2\% & 23.7\% & 23.1\%
& \texttt{layout} & NVIDIA & 85.5\% & 14.4\% & 0.1\% \\
& Intel & 79.2\% & 7.8\% & 13.0\%
& & Intel & 97.6\% & 1.3\% & 1.1\% \\
& AMD & 5.9\% & 1.1\% & 93.1\%
& & AMD & 4.5\% & 84.6\% & 10.9\% \\
\hline
\texttt{randomAccess} & NVIDIA & 45.8\% & 33.4\% & 20.7\%
& \texttt{colorwheel} & NVIDIA & 0.0\% & 87.3\% & 12.7\% \\
& Intel & 79.0\% & 14.1\% & 6.8\%
& & Intel & 68.8\% & 24.3\% & 6.9\% \\
& AMD & 9.0\% & 86.8\% & 4.2\%
& & AMD & 17.1\% & 0.3\% & 82.6\% \\
\hline
\end{tabular}
\caption{Outcome ratios across the aligned three-backend benchmark set. Percentages are rounded.}
\label{tab:backend-coarse-summary}
\end{table*}

Outcome distributions differ consistently and substantially by 
backend across all benchmarks, with clear shifts in the balance 
between \textsc{Masked}, \textsc{SDC}, and \textsc{Failure} outcomes.
The largest separations appear in \texttt{dense-embedding}, \texttt{layout}, \texttt{colorwheel}, and \texttt{jacobi}, where backend choice strongly influences the dominant outcome class. Even in cases where coarse outcome ratios appear similar, the manifestation of non-masked faults differs across platforms.

Cross-backend agreement is limited: only 10.4\% of aligned injections produce the same outcome across all three platforms, while most cases exhibit partial or complete disagreement. Interestingly, pairwise transitions show consistent divergence. For example, outcomes that are \textsc{Masked} on Intel or NVIDIA frequently become \textsc{SDC} on AMD (over 83\% of cases), while \textsc{SDC} on NVIDIA is masked by Intel in more than half of cases. AMD also shifts outcomes toward higher severity more often than either NVIDIA or Intel.

\textbf{Finding 1.} Semantically aligned IR-level faults do not yield backend-stable resilience behavior. The same perturbation may be masked on one backend, silently corrupt output on another, and surface as a hard failure on a third, so resilience conclusions drawn from a vendor do not transfer to others.

\subsection{Site Class and Value Type Analysis}
\label{sec:siteclass}

\begin{qbox}
\textbf{Question 2.} How does fault outcome vary across injection site classes and value type (§\ref{sec:IR-FI}), and does this sensitivity differ across GPU backends?
\end{qbox}

\begin{figure}[t]
    \centering
    \includegraphics[width=0.87\linewidth]{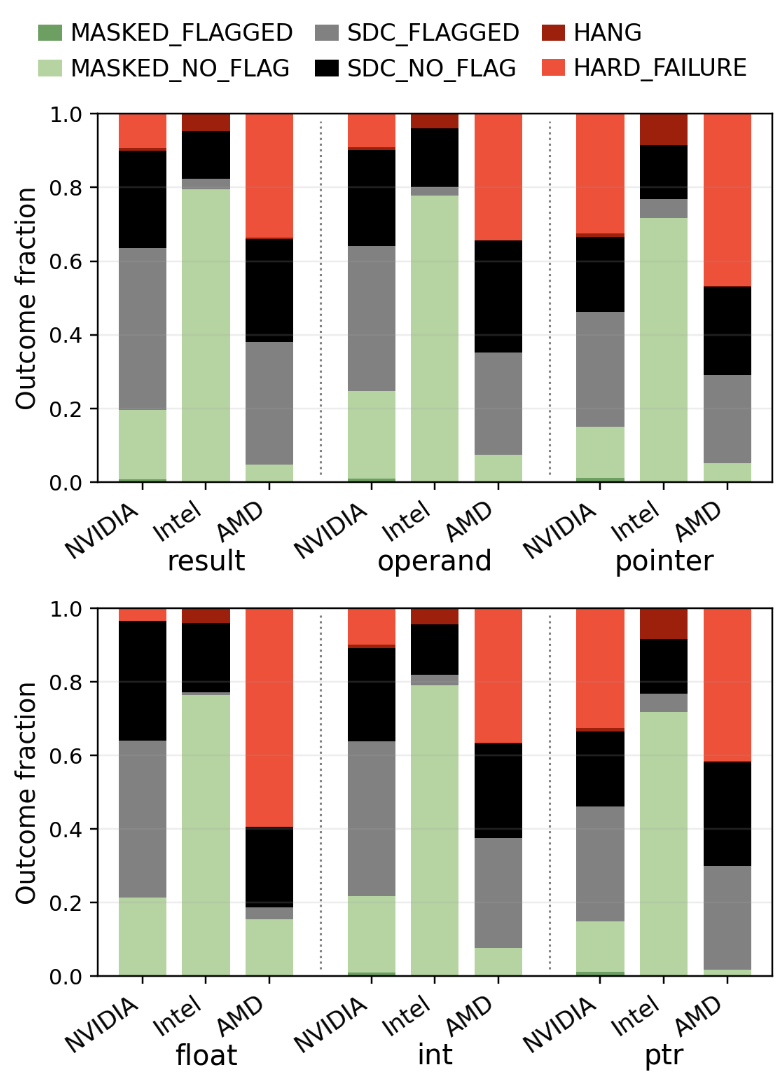}
    \caption{Fine-grained outcome distributions broken down by injection class (top) and value type (bottom).}
    \label{fig:siteclass-typekind}
\end{figure}

\begin{figure*}[t]
    \centering
    \includegraphics[width=\linewidth]{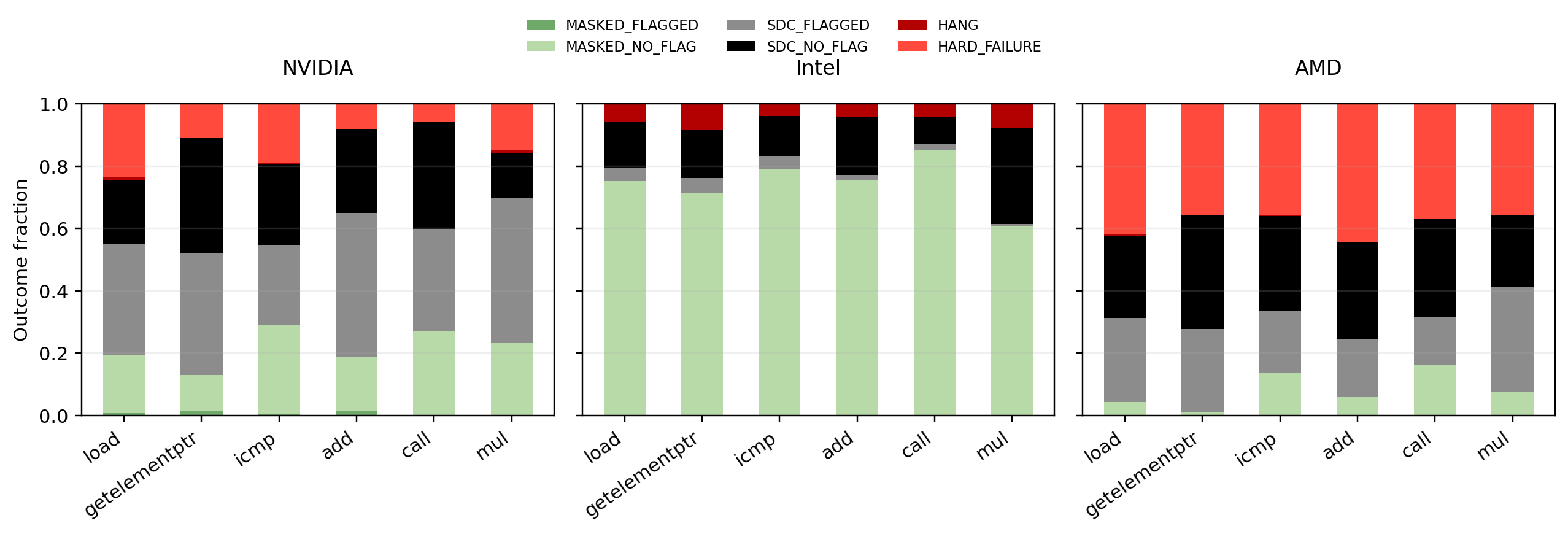}
    \caption{Fine-grained outcome distributions by the six most frequently injected LLVM IR instructions. }
    \label{fig:opcode-global-signature}
\end{figure*}

Figure~\ref{fig:siteclass-typekind} shows the breakdown by 
site class and value type, while 
Figure~\ref{fig:opcode-global-signature} 
shows which instruction-level structures sharpen these backend differences. Together, these views 
confirm that outcome sensitivity is strongly shaped by site class and value type, 
but not in the same way across all backends.

\textit{Injection class.} 
The data show that site class matters, but differently on each backend. 
Across \emph{result} and \emph{operand} sites, Intel remains 
predominantly \textsc{Masked}, NVIDIA is \textsc{SDC}-heavy, and AMD 
is largely non-masked. \emph{Pointer} sites are the strongest 
separator:NVIDIA shows a mix of \textsc{SDC} and \textsc{Failure}, 
Intel remains mostly \textsc{Masked} but with its highest 
\textsc{Failure} and \textsc{Hang} rates among all site classes, and 
AMD produces minimal masking with the highest non-masked fraction 
across all three classes.

\textit{Value type.}
Value type shows a consistent pattern across backends, with 
floating-point sites have the highest masking rates overall, while 
integer and pointer sites concentrate \textsc{Failure} and 
\textsc{SDC\_No\_Flag} outcomes, as perturbations in control-flow, 
indexing, and address computation propagate more directly to 
observable failures than small numerical perturbations in iterative 
arithmetic.

\textit{Opcode-level structure.}
At the opcode level, divergence is concentrated in memory and 
address-generation instruction families 
(Figure~\ref{fig:opcode-global-signature}). Integer and 
control-bearing opcodes such as \texttt{xor}, \texttt{select}, 
\texttt{shl}, and \texttt{udiv} show particularly strong 
cross-backend separation. They are nearly entirely \textsc{Masked} 
on Intel but produce predominantly \textsc{SDC} on NVIDIA and 
\textsc{Failure} on AMD, confirming that divergence is not limited 
to memory operations.

\textbf{Finding 2.} Fault outcome is strongly structured by 
injection class and value type, but the sensitivity is 
backend-dependent. Pointer-bearing and integer/control-bearing sites drive the strongest cross-backend divergence. Intel masks heavily 
across all classes, NVIDIA concentrates non-masked faults as 
\textsc{SDC}, and AMD shifts aggressively toward hard failure, 
particularly at pointer and integer sites.

\subsection{Workload Analysis}
\label{sec:results-workload}

\begin{qbox}
\textbf{Question 3.} To what extent does workload structure determine resilience behavior, and are certain computational motifs systematically more vulnerable than others across all three platforms?
\end{qbox}

\begin{figure*}[t]
    \centering
    \includegraphics[width=0.85\linewidth]{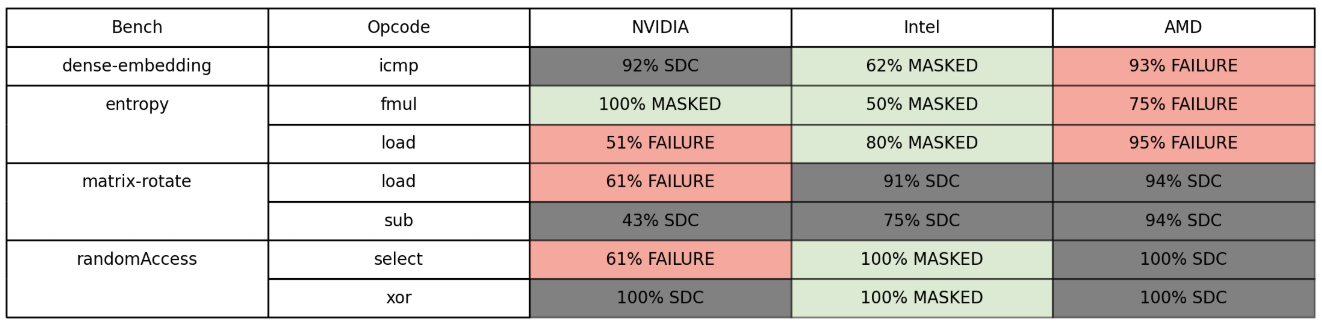}
    \caption{Representative backend divergence points: benchmark-opcode pairs where the dominant coarse outcome class differs across NVIDIA, Intel, and AMD. Cell color indicates the dominant outcome: green (\textsc{Masked}), gray (\textsc{SDC}), and red (\textsc{Failure}). Percentages reflect the fraction of injections producing the dominant outcome for that backend.}
    \label{fig:backend-flip-points}
\end{figure*}

Table~\ref{tab:backend-coarse-summary} provides the benchmark-level coarse outcome ratios, and Figure~\ref{fig:backend-flip-points} highlights representative benchmark-opcode combinations where the dominant outcome class differs across vendors. Together, they show that workload structure exposes different backend-specific regimes rather than preserving a single global resilience ordering.

Workload structure exposes different backend-specific regimes rather than preserving a single global resilience ordering. Iterative kernels such as \texttt{jacobi} and \texttt{pathfinder} show a consistent split: Intel is predominantly masked, NVIDIA is split between masked and SDC, and AMD is strongly SDC-heavy. Memory-bound and transformation-heavy kernels exhibit different patterns, with some workloads showing SDC-dominant behavior across Intel and AMD while others drive AMD toward failure-heavy regimes. In several cases, the dominant outcome class differs entirely across vendors for the same benchmark--opcode combination, confirming that fault manifestation depends jointly on workload context and backend execution behavior.

These patterns are consistent with documented backend execution behavior. Intel's SYCL execution model exposes sub-group and work-group level execution and synchronization semantics that can influence how perturbations propagate through collective and iterative kernels~\cite{IntelOneAPI}. NVIDIA's CUDA programming model specifies that illegal or out-of-bounds memory accesses may result in runtime errors or program termination, which is consistent with the larger fraction of overt failures observed in address-sensitive paths~\cite{CUDAProgGuide}. Similarly, AMD's HIP programming model permits execution to continue when memory accesses remain within valid regions, while invalid accesses may produce incorrect results or runtime faults, consistent with the observed mixture of \textsc{SDC}-heavy and failure-heavy regimes~\cite{AMDHIP}.

\textbf{Finding 3.} Workload structure does not produce a single 
backend resilience ranking. Different computational motifs expose 
different vendor-specific propagation regimes, so resilience is 
workload-aware as well as backend-aware. The same instruction class 
can be fully \textsc{Masked} on one backend, produce silent corruption 
on another, and trigger hard failure on a third depending on the 
workload in which it appears.

\subsection{SDC and Hang Analysis}
\label{sec:results-indepth}

\begin{qbox}
\textbf{Question 4.} Where is silent data corruption risk concentrated across backends, workloads, and injection sites, and what distinguishes Intel hang behavior from the hard-failure modes observed on NVIDIA and AMD?
\end{qbox}

To ground the cross-architecture patterns identified in Questions 
1--3, we examine two forms of non-masked backend divergence that 
appear repeatedly across the study: \emph{silent data corruption 
risk} and \emph{hang behavior}.

\textit{Silent Corruption Risk.}
Silent data corruption is the most operationally dangerous outcome 
because the application completes while producing incorrect output. 
Our data show that \textsc{SDC} risk is not uniformly distributed 
across backends; it is concentrated in specific combinations of 
backend, workload, and site structure. Representative examples include 
AMD pointer and operand sites in \texttt{jacobi} and 
\texttt{matrix-rotate}, Intel pointer sites in \texttt{matrix-rotate}, 
and NVIDIA operand and result sites in \texttt{dense-embedding}, all of 
which are strongly \textsc{SDC}-heavy.

At the aggregate level, \textsc{SDC\_No\_Flag} accounts for 16.3\% 
of aligned injections on NVIDIA, 14.5\% on Intel, and 12.5\% on 
AMD. Structurally, \textsc{SDC} risk is concentrated at pointer-bearing 
and operand sites, especially in workloads dominated by index remapping, 
iterative propagation, or dense arithmetic reuse. Memory- and 
address-related opcodes such as \texttt{load}, \texttt{store}, and 
\texttt{getelementptr} are major contributors, because corrupted 
addresses can still refer to legal but incorrect locations. Integer and 
control-bearing instructions such as \texttt{xor}, \texttt{select}, 
\texttt{shl}, and \texttt{udiv} also contribute strongly on NVIDIA and 
AMD while remaining much more often masked on Intel. These trends show 
that \textsc{SDC} risk is highly structured rather than uniform across 
backends, and that the same backend can be \textsc{SDC}-heavy on one 
workload and failure-heavy on another.

\textit{Hang Behavior.}
Hangs are operationally severe because they stall progress and require a timeout or external intervention. In this study, hang behavior is 
overwhelmingly concentrated on Intel. Across the aligned benchmark set, 
Intel produces 7.1\% \textsc{Hang}, compared with 0.5\% on NVIDIA and 
0.2\% on AMD. Hang behavior on Intel is workload and site class sensitive. It is the highest in 
\texttt{entropy} (13.0\%), followed by \texttt{jacobi} (7.8\%), 
\texttt{pathfinder} (7.7\%), \texttt{matrix-rotate} (7.2\%), 
\texttt{colorwheel} (6.9\%), and \texttt{randomAccess} (6.8\%). By site class, pointer injections show the highest hang share on 
Intel; address- and pointer-related combinations are especially 
hang-prone, particularly \texttt{addrspacecast}, 
\texttt{getelementptr}, and \texttt{load} at pointer sites. 

\textbf{Finding 4.} \textsc{SDC} risk is concentrated in 
structured backend--workload--site combinations where perturbed 
values remain legally executable while producing incorrect output. 
Hang risk is an Intel-dominated mode driven by pointer-bearing and 
address-related sites, where non-masked faults retain forward 
progress long enough to stall rather than terminate. Practical 
mitigation must therefore be backend-aware. Output validation for 
\textsc{SDC}, timeout-aware stall recovery for Intel hangs, and 
fast failure detection for NVIDIA and AMD hard failures.

\subsection{Case Study: \texttt{matrix-rotate}}
\label{sec:results-casestudies}

\begin{qbox}
\textbf{Question 5.} What does \texttt{matrix-rotate} reveal 
about the backend-specific mechanisms that determine whether an aligned IR-level fault produces a completed execution with corrupted output or a hard failure, and why does a completed execution alone not imply correctness for this 
workload?
\end{qbox}

\begin{figure*}[t]
    \centering
    \includegraphics[width=\linewidth]{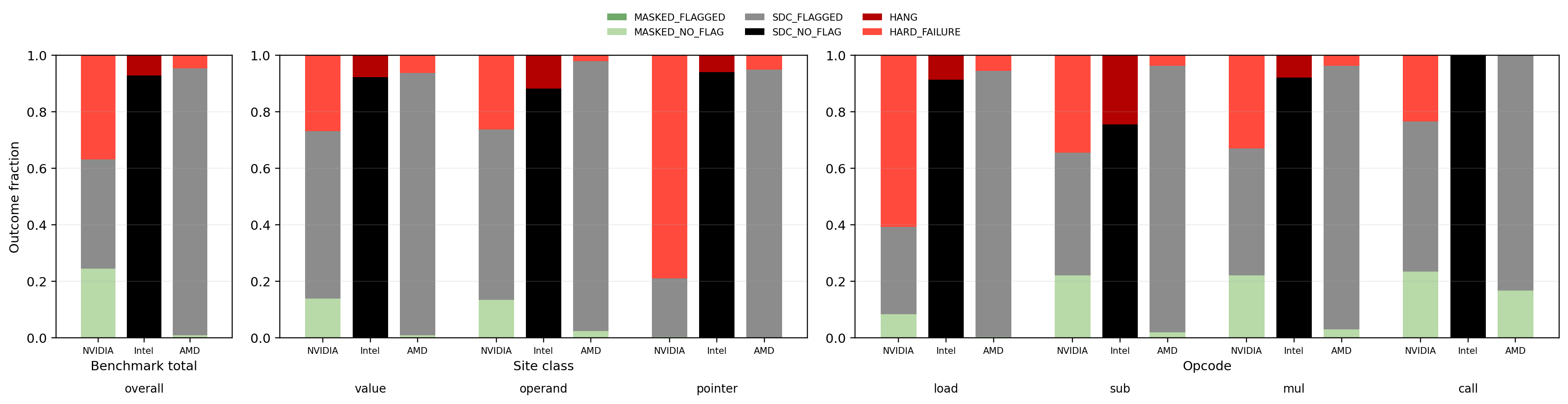}
    \caption{\texttt{matrix-rotate} microstudy: fine-grained outcome 
    distributions by overall benchmark total, site class (value, 
    operand, pointer), and top opcodes (\texttt{load}, \texttt{sub}, 
    \texttt{mul}, \texttt{call}) across NVIDIA, Intel, and AMD.}
    \label{fig:microstudy-matrix-rotate}
\end{figure*}

Figure~\ref{fig:microstudy-matrix-rotate} presents a detailed 
 breakdown of \texttt{matrix-rotate} across the 
fine-grained taxonomy, site-class distributions, and top divergent 
opcodes. We focus on this benchmark because it exemplifies the most 
operationally subtle form of backend divergence: completed 
corruption on Intel and AMD versus hard failure on NVIDIA.

At the coarse level, Intel and AMD are almost entirely 
\textsc{SDC}-heavy (92.8\% and 94.4\% respectively) while NVIDIA 
shows a substantially larger failure component at 36.9\%. This benchmark 
is dominated by index remapping and in-place data movement, so many 
perturbed values lead to legal execution and in-bounds memory access 
even when the final output is incorrect. Completion is therefore not 
evidence of resilience.

The pointer-site view sharpens this picture most clearly. Pointer 
injections in \texttt{matrix-rotate} produce 79.1\% \textsc{Failure} 
on NVIDIA, but 94.0\% \textsc{SDC} on Intel and 94.8\% \textsc{SDC} 
on AMD, one of the strongest benchmark-specific examples of 
backend-dependent manifestation in the study. The same aligned class 
of faults pushes NVIDIA toward overt failure while leaving Intel and 
AMD in a completed-yet-incorrect regime, consistent with the kernel's 
dominant operations being address and index transformations whose 
perturbed values frequently remain in-bounds while redirecting data 
to the wrong location. The opcode-level view in 
Figure~\ref{fig:microstudy-matrix-rotate} confirms that this pattern 
holds across \texttt{load}, \texttt{sub}, \texttt{mul}, and 
\texttt{call} instructions, none of which escape the same 
NVIDIA-failure versus Intel/AMD-SDC split.

\textbf{Finding 5.} In \texttt{matrix-rotate}, the same aligned pointer-site fault 
produces 79.1\% \textsc{Failure} on NVIDIA but over 94\% 
\textsc{SDC} on Intel and AMD, driven by the kernel's index and 
address transformations that frequently yield perturbed values 
remaining legally in-bounds. For this class of workload, output validation against a golden reference is the most reliable safeguard.

\section{Implications for GPU Resilience Practice}
\label{sec:implications}

The observed backend-specific fault behaviors have direct implications for how resilience mechanisms should be deployed in practice. Rather than applying uniform strategies across platforms, mitigation should reflect the dominant failure modes observed for each backend.

\textbf{NVIDIA.} Faults more frequently surface as explicit runtime failures, particularly in address-sensitive execution paths and workloads with irregular memory access or index remapping. This behavior favors mitigation centered on rapid detection and controlled recovery, including early termination, checkpoint/restart, and failure-aware execution models~\cite{Moody2010}. Because these faults are more likely to terminate execution overtly, recovery mechanisms are especially relevant when targeting NVIDIA backends.

\textbf{AMD.} Faults are substantially more likely to produce corrupted outputs than explicit failures, making silent data corruption a primary concern. Mitigation on AMD platforms should therefore emphasize application-level output validation and error detection. Algorithm-based fault tolerance (ABFT)~\cite{Huang1984,Chen2013} and redundant computation are particularly relevant because they can detect silent errors without relying on runtime failure signals. This is especially important for iterative and transformation-heavy workloads, where perturbed values frequently remain in-bounds and propagate to incorrect outputs without triggering hard failures.

\textbf{Intel.} Hang behavior is substantially more common on Intel, accounting for 7.1\% of injections compared with 0.5\% on NVIDIA and 0.2\% on AMD. Faults that escape masking can retain forward progress long enough to stall rather than terminate cleanly, making timeout enforcement, hang detection, stall recovery, and failure-aware scheduling important mitigation concerns. Systems targeting Intel backends should therefore pair these mechanisms with checkpoint/restart where recovery from stalled execution is required~\cite{Moody2010}.

These mitigation mechanisms introduce different costs and must be configured according to workload characteristics. Checkpoint/restart introduces storage and synchronization overhead, with appropriate checkpoint granularity depending on application state size, checkpoint cost, and expected failure frequency. ABFT, output validation, and redundant computation introduce additional arithmetic, validation, or duplicated execution cost. Our results do not quantify these overheads or prescribe fixed checkpoint intervals; instead, they identify the backend--workload combinations in which particular forms of protection are most warranted.

\textbf{Practical Impact.} During large-scale campaign execution on Aurora, segmentation faults caused nodes to go offline and trigger system maintenance, prompting direct contact from the ALCF operations team. Investigation revealed that these hangs exposed a previously undetected issue in Intel's GPU system stack. The ALCF operations team subsequently debugged the issue and deployed a fix on Aurora. All Intel results reported in this paper were collected after this fix was applied, so the defect does not affect the reported outcome distributions. This incident demonstrates that cross-architecture fault injection at the IR level can expose system-level vulnerabilities in production GPU environments and, in this case, contributed to identifying and correcting an issue in the deployed system stack.

\section{Related Work}
\label{sec:related}

Prior work on GPU resilience spans architectural, microarchitectural, software, and compiler-level perspectives. Foundational architectural studies establish the fault-outcome concepts and evaluation principles underlying resilience analysis~\cite{MukherjeeHPCA2005,ReisTACO2005}. On GPUs, early vulnerability studies show that fault manifestation depends strongly on both microarchitectural location and application behavior. Tan et al.~\cite{Tan2011} analyze soft-error vulnerability in GPGPU microarchitecture, while Vallero et al.~\cite{Vallero2018} compare reliability behavior across NVIDIA and AMD GPUs. More broadly, prior evidence comes from both modeled or simulation-based analyses and field studies of deployed systems. Radiation-induced and large-scale system studies~\cite{Oliveira2016,TiwariHPCA2015,Sridharan2015,DiMartino2014} reinforce the same conclusion that GPU fault outcomes vary substantially with hardware structure, workload, and deployment environment. Together, these studies establish why GPU resilience must be studied empirically, but they do not provide a portable mechanism for injecting semantically aligned faults and observing their manifestation through real application runs across vendor ecosystems.

The methods most directly related to BitIR begins with software-visible GPU fault injection. GPU-Qin~\cite{Fang2014} is among the earliest frameworks for systematic GPU resilience evaluation through software-level fault injection. Later post-backend frameworks improve realism by injecting faults into vendor-specific instruction streams after compilation. SASSIFI~\cite{SASSIFI2017} operates at the NVIDIA SASS level, NVBitFI~\cite{NVBitFI2021} extends this model with dynamic binary instrumentation, and gpuFI-4~\cite{gpuFI42022} further refines campaign design and site classification. These systems are effective for real-hardware studies, but their reliance on backend-specific representations prevents natural alignment of equivalent faults across CUDA, HIP, and SYCL toolchains. Recent cross-layer studies~\cite{DemystifyingGPU2021,YangISPASS2024} make this limitation even clearer by showing that resilience conclusions can shift materially with the abstraction level and methodology used for injection.

A second line of work similar to BitIR provides the compiler and application level foundation for our approach. LLFI~\cite{LLFI2015}, FlipIt~\cite{FlipIt2014}, and REFINE~\cite{REFINE2017} establish the viability of IR-level fault injection for CPU-based HPC workloads, showing that LLVM IR can support both portability and meaningful propagation analysis. GPU-oriented extensions move this reasoning closer to our setting. LLFI-GPU~\cite{LLFIGPU} brings LLVM based FI to GPU applications, Li et al.~\cite{LiSC2016} study software-visible GPU error propagation, and GPU-Trident~\cite{GPUTrident2020} develops efficient models of GPU error propagation. Related application- and compiler-level studies~\cite{PRISM2018,Kalra2020,CompilerOptGPU2024} further show that program structure, data representation, compiler transformation, and execution context can significantly affect resilience. These works justify LLVM IR as a meaningful analysis layer, but none establishes a unified framework that uses aligned LLVM IR fault sites to drive directly comparable campaigns across NVIDIA, AMD, and Intel GPUs on production systems.

Many prior GPU fault injection studies rely on statistical or randomized sampling of fault locations for tractability~\cite{Fang2014,SASSIFI2017,NVBitFI2021,gpuFI42022}, which is a practical design choice for large campaigns but can miss structured divergence concentrated in specific opcodes, site classes, or workload contexts. For cross-vendor comparison, this limitation is especially important. If aligned faults are sampled sparsely, backend-specific differences may be underrepresented or missed entirely. BitIR evaluates the complete aligned injection space for each retained benchmark rather than doing any sampling, preserving fine-grained comparability across vendors.

Our outcome classification follows the standard masked--SDC--failure taxonomy established in architectural resilience studies~\cite{MukherjeeHPCA2005,ReisTACO2005}, with refinements for GPU workloads following Kalra et al.~\cite{Kalra2020}. Unlike prior approaches that rely on application-reported status, BitIR classifies all outcomes against golden references, enabling consistent interpretation across vendor runtime systems. We additionally distinguish \textsc{HANG} from \textsc{HARD\_FAILURE}, which proves important in cross-vendor analysis.

Taken together, prior work establishes the importance of GPU resilience analysis~\cite{Tan2011,Oliveira2016,TiwariHPCA2015,Vallero2018}, 
the value of real-hardware fault injection~\cite{Fang2014,SASSIFI2017,NVBitFI2021,gpuFI42022,DemystifyingGPU2021}, 
and the usefulness of higher-level program representations for studying fault propagation~\cite{LLFI2015,FlipIt2014,REFINE2017,LiSC2016,Kalra2020,CompilerOptGPU2024}. 
These directions build on broader architectural and system-level reliability studies~\cite{MukherjeeHPCA2005,ReisTACO2005,Sridharan2015,DiMartino2014}. What remains missing is a system that simultaneously provides cross architecture fault modeling at the LLVM IR level, execution on real production GPU hardware, and direct cross-vendor comparability through aligned injection sites. BitIR is designed to fill that gap.

\section{Conclusion}

We presented BitIR, an LLVM IR-based fault injection framework for 
cross-architecture resilience analysis of heterogeneous GPU applications. 
By injecting semantically aligned single-bit faults before backend 
lowering, BitIR enables direct comparison of software-visible fault behavior across NVIDIA, Intel, and AMD GPUs on three leadership-class 
supercomputers under a shared fault model. 
Our campaigns show that resilience is not backend-invariant. The same aligned fault can be 
masked on one platform, manifest as silent data corruption on 
another, and surface as overt failure on a third. These differences 
are especially pronounced for memory-access, address-generation, 
pointer-bearing, and integer/control-bearing sites, and vary 
substantially by workload, underscoring the need for backend-aware 
and workload-aware fault mitigation.

Our results also have direct implications for resilience practice. 
Backend-specific behaviors require different mitigation priorities: rapid recovery for NVIDIA, output validation for AMD, and 
timeout enforcement with hang detection for Intel, as well as workload 
structure compounds these differences. Our campaigns on Aurora further surface latent system-level vulnerabilities; a previously undetected issue in Intel's GPU software stack is identified with the ALCF operations team, who subsequently deployed a fix on Aurora.
The implementation and datasets will be released as open source on 
GitHub upon 
acceptance.\footnote{\url{https://github.com/SPEAR-UIC/BitIR}}

Future work will extend BitIR to multi-bit and burst-error fault 
scenarios, and apply the cross-backend patterns identified here to 
guide backend-aware mitigation design. 
We also plan to extend the evaluation to multi-kernel and full application workloads to determine how larger program state and deeper execution structure affect cross-backend masking and propagation. \\

\section*{Acknowledgment}
We acknowledge the use of computational resources provided by the Argonne Leadership Computing Facility (ALCF), the Oak Ridge Leadership Computing Facility (OLCF), and other supporting computing facilities. 

This work was supported in part by the U.S. National Science Foundation under Grants Nos. OAC-2402901 and CCF-2515009 and in part by the U.S. Department of Energy under Contract No. DE-SC0024271. We also acknowledge support from the U.S. Department of Energy, Office of Science, under Grant No. DE-SC0023524 through the Chicagoland Computational Traineeship in High Energy Particle Physics Fellowship.

\bibliographystyle{IEEEtran}
\bibliography{GPU-FI/SC_technical_paper/references}
\end{document}